\documentclass[sigconf,authorversion=true,natbib=true,balance=true]{acmart}

\usepackage{amsmath}%
\usepackage{rotating}%
\usepackage{color,soul}
\usepackage{tabu}%
\usepackage{multirow}%
\usepackage{acronym}
\usepackage{makecell}
\usepackage{enumitem}
\usepackage{listings} % For code listing
\usepackage{xcolor}
\usepackage{array}

\definecolor{myyellow}{rgb}{1.0, 1.0, 0.5}
\definecolor{myblue}{rgb}{0.7, 0.9, 1.0}
\definecolor{myorange}{rgb}{1.0, 0.8, 0.5}
\definecolor{mypurple}{rgb}{0.9, 0.8, 1.0}
\definecolor{mygray}{rgb}{0.9, 0.9, 0.9}

\definecolor{mylight}{rgb}{0.90, 0.95, 1.0}

\setlist{parsep=.1em,itemsep=.1em,topsep=.1em,leftmargin=1em}

\newcounter{UseArxiv}
\newcommand{\car}{TREC CAR Y3}
\newcommand{\tqa}{TQA}
\newcommand{\dl}{TREC DL}

\newcommand{\dlsecond}{TREC DL 2020}

\newcommand{\questionanswering}{quest\-ion ans\-wer\-ing}

\ifnum\value{UseArxiv}=1

\newcommand{\system}{Autograding}
\newcommand{\EXAM}{Autograding}
\newcommand{\thesystem}{the Autograding Workbench}

\newcommand{\oursystem}{our Autograding Workbench}
\newcommand{\Oursystem}{Our Autograding Workbench}
\newcommand{\Ourapproach}{Our Autograding approach}
\newcommand{\theapproach}{the Autograding approach}
\newcommand{\examcover}{Autograde Cover}
\newcommand{\examqrels}{Autograde Qrels}
\newcommand{\systemfile}{autograde}

\newcommand{\resourceurl}[0]{https://github.com/TREMA-UNH/autograding-workbench}

\else

\newcommand{\system}{Rubric}

\newcommand{\EXAM}{RUBRIC}
\newcommand{\thesystem}{the Rubric-Grading Workbench}

\newcommand{\oursystem}{our Rubric-Grading Workbench}
\newcommand{\Oursystem}{Our Rubric-Grading Workbench}
\newcommand{\Ourapproach}{Our Rubric-Grading approach}
\newcommand{\theapproach}{the Rubric-Grading approach}
\newcommand{\examcover}{Rubric-Cover}
\newcommand{\examqrels}{Rubric-Qrels}
\newcommand{\systemfile}{rubric}

\newcommand{\resourceurl}[0]{https://github.com/TREMA-UNH/rubric-grading-workbench}

\fi

\newcommand{\treceval}{\texttt{trec\_eval}}
\newcommand{\squad}{SQuAD2}

\newcommand{\commentout}[1]{}

\newlist{compactdesc}{description}{3}
\setlist[compactdesc]{topsep=0pt,partopsep=0pt,itemsep=0pt,parsep=0pt}

\ifnum\value{UseArxiv}=1

\acmConference[SIGIR '24]{Proceedings of the 47th International ACM SIGIR
Conference on Research and Development in Information Retrieval}{July 14--18,
2024}{Washington, DC, USA}
\else

\copyrightyear{2024}
\acmYear{2024}

\setcopyright{rightsretained}
\acmConference[SIGIR '24]{Proceedings of the 47th International ACM SIGIR Conference on Research and Development in Information Retrieval}{July 14--18, 2024}{Washington, DC, USA} \acmBooktitle{Proceedings of the 47th International ACM SIGIR Conference on Research and Development in Information Retrieval (SIGIR '24), July 14--18, 2024, Washington, DC, USA} \acmDOI{10.1145/3626772.3657871} \acmISBN{979-8-4007-0431-4/24/07}

\fi 
\makeatletter
\gdef\@copyrightpermission{
   \begin{minipage}{0.3\columnwidth}
     \href{https://creativecommons.org/licenses/by-nc-sa/4.0/}{\includegraphics[width=0.90\textwidth]{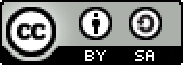}}
   \end{minipage}\hfill
   \begin{minipage}{0.7\columnwidth}
     \href{https://creativecommons.org/licenses/by-nc-sa/4.0/}{This work is licensed under a Creative Commons Attribution-ShareAlike International 4.0 License.}
   \end{minipage}
   \vspace{5pt}
}
\makeatother

\begin{document}

\ifnum\value{UseArxiv}=1
\title{A Workbench for Autograding Retrieve/Generate Systems}
\else
\title{A Workbench for Autograding Retrieve/Generate Systems}
\fi
%\title{EXAM++: Exam Answerability Metric with LLMs}
% \title{EXAM++: Evaluating with Exam Questions instead of Relevance Judgments}

\author{Laura Dietz}
\email{dietz@cs.unh.edu}
% \orcid{\ld{needed!}}
\orcid{0000-0003-1624-3907}
\affiliation{%
  \institution{University of New Hampshire}
  \city{Durham, NH}
  \country{USA}
}

% \ifnum\value{UseArxiv}=1

% \begin{abstract}
% This resource paper addresses the challenge of evaluating Information Retrieval (IR) systems in the era of autoregressive Large Language Models (LLMs). Traditional methods relying on passage-level judgments are no longer effective due to the diversity of responses generated by LLM-based systems. We provide a workbench to explore several alternative evaluation approaches to judge the relevance of a system's response that incorporate LLMs: 1. Asking an LLM whether the response is relevant; 2. Asking the LLM which set of nuggets (i.e., relevant key facts) is covered in the response; 3. Asking the LLM to answer a set of exam questions with the response.

% This workbench aims to facilitate the development of new, re\-usa\-ble test collections. Researchers can manually refine sets of nug\-gets and exam questions, observing their impact on system evaluation and leaderboard rankings.\footnote{Resource available at \url{ \resourceurl }}

% \end{abstract}

% \else

\begin{abstract}

This resource paper addresses the challenge of evaluating Information Retrieval (IR) systems in the era of autoregressive Large Language Models (LLMs). Traditional methods relying on passage-level judgments are no longer effective due to the diversity of responses generated by LLM-based systems. We provide a workbench to explore several alternative evaluation approaches to judge the relevance of a system's response that incorporate LLMs: 1. Asking an LLM whether the response is relevant; 2. Asking the LLM which set of nuggets (i.e., relevant key facts) is covered in the response; 3. Asking the LLM to answer a set of exam questions with the response.

This workbench aims to facilitate the development of new, re\-usa\-ble test collections. Researchers can manually refine sets of nug\-gets and exam questions, observing their impact on system evaluation and leaderboard rankings.\footnote{Resource available at \url{ \resourceurl } \\ 
\ifnum\value{UseArxiv}=1
To appear in the Resource \& Reproducibility Track of SIGIR 2024.
\fi
}

\end{abstract}

% \fi




\maketitle

\acrodef{LLM}[LLM]{Large Language Model}

\section{Introduction}
\label{sec:introduction}

\sloppypar The evaluation of IR systems is traditionally based on passage-level relevance judgments. The assumption is that any good system should retrieve the set of as-relevant judged passages. Once collected, such judgments can be reused in the form of a test collection. However, with the advent of strong auto-regressive LLMs, such as ChatGPT, Llama2, Mistral, and FLAN-T5, the research community is developing new IR systems to which the traditional approach is no longer applicable. The problem is that even for the same search query, such systems will produce a plethora of different responses, each slightly different in linguistic style, level of detail, and tone. Even when asking the same system twice, two slightly different responses will be obtained. Hence, we do not expect to see the exact same system response more than once.  While such system responses can still be evaluated by manually judging the relevance of the response, the resulting test collection will not be re-usable under the traditional evaluation paradigm.

There are some ways out of this conundrum:

\begin{enumerate}
    \item \textbf{Hire more judges or perform A/B tests.}---While this is possible in industry, it is too costly for academic use and does not support reproducible research.
    \item \textbf{Use a summarization evaluation metric}. Measure the similarity of system responses to the known correct responses.---While this has demonstrated success \citep{passage-rouge,zhang2019bertscore} the match-quality deteriorates when longer system responses are generated.
    \item \textbf{Use Large Language Models (LLMs) to replace human jud\-ges}---While \citet{thomas2023large,faggioli2023perspectives,sun2023chatgpt} empirically demonstrated success, there are several concerns ab\-out the trustworthiness of LLMs, especially in a 100\% automatic evaluation approach \cite{faggioli2023perspectives,faggioli2024determines}.
    \item \textbf{Let human judges define which pieces of information are relevant, use automatic alignment methods.} This approach is used in nugget-based evaluation \cite{lin2006will,smucker2008qanugget,pavlu2012ir,sakai2011click} and the EXAM Answerability Metric \citep{sander2021exam}. Our resource builds on these ideas.
\end{enumerate}

\paragraph{Contributions.}
In this resource paper, we provide a workbench that builds on three ideas for LLM-based ``autograding'' approaches: 
\begin{description}
\item [Question:]  Relevant responses answer exam questions \citep{sander2021exam,farzi2024exam},
\item [Nugget:] Relevant responses mention key nuggets \citep{lin2006will,pavlu2012ir}, 
\item [Direct:] Relevance can be directly predicted \citep{thomas2023large,faggioli2023perspectives,sun2023chatgpt,liang2022holistic}.
\end{description}

\noindent We are releasing \thesystem{} software, along with usage examples based on \dlsecond{}. (URL in abstract.)

\begin{figure*}
\ifnum\value{UseArxiv}=1
    \includegraphics[width=0.78\textwidth]{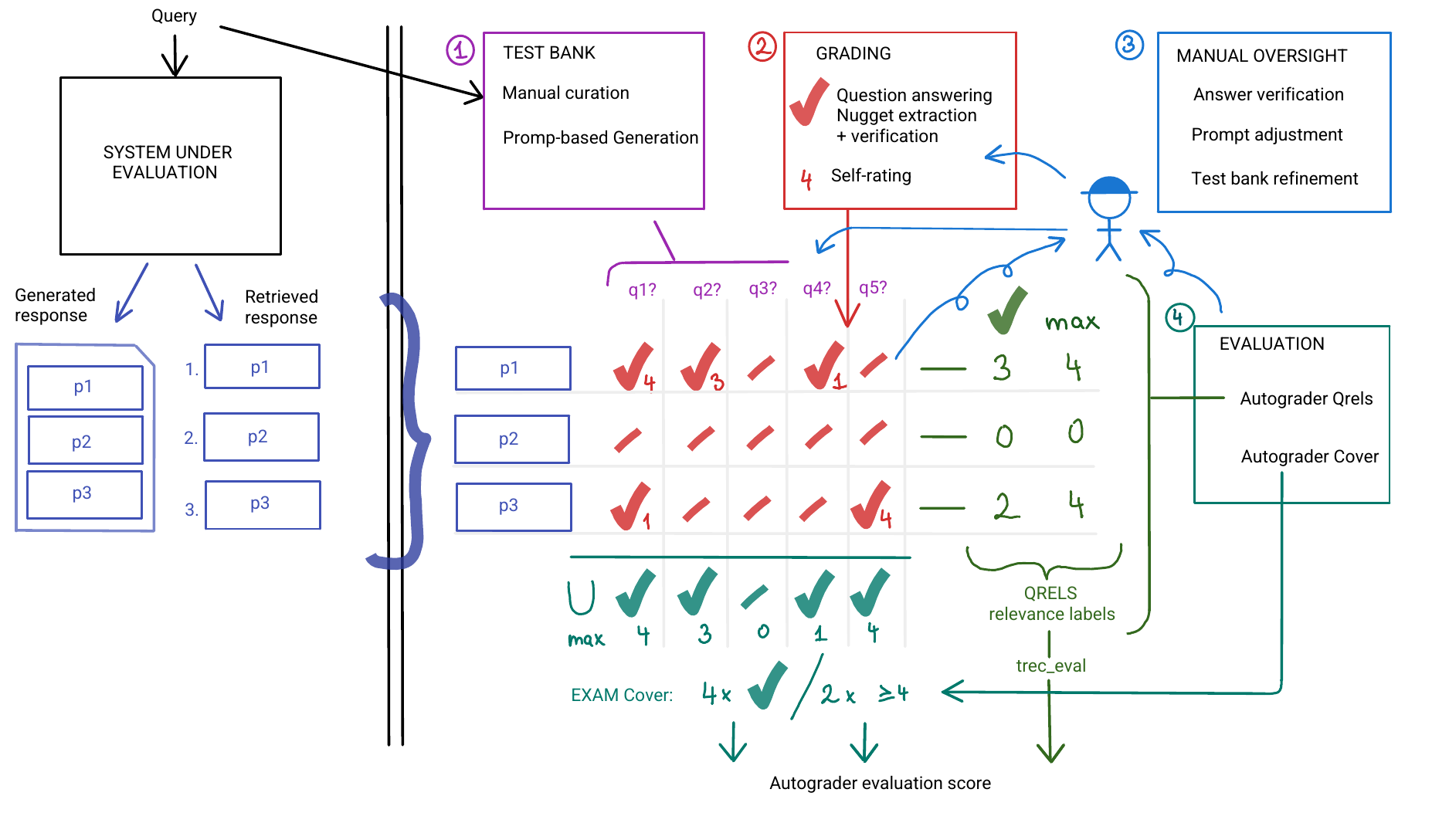}
    \caption{\system{} Process: Phase 1: Test bank creation with semi-automatic methods. Phase 2: Automatic grading with prompt-based LLMs. Phase 3: Manual verification an oversight. Phase 4: Evaluation via \treceval{} or \examcover{}. Results from any phase are used by the human-in-the-loop to refine the test bank and adjust prompts to ensure that the automatic grading agrees with the human understanding of relevance.} \label{fig:approach}
\else 
    \includegraphics[width=0.9\textwidth]{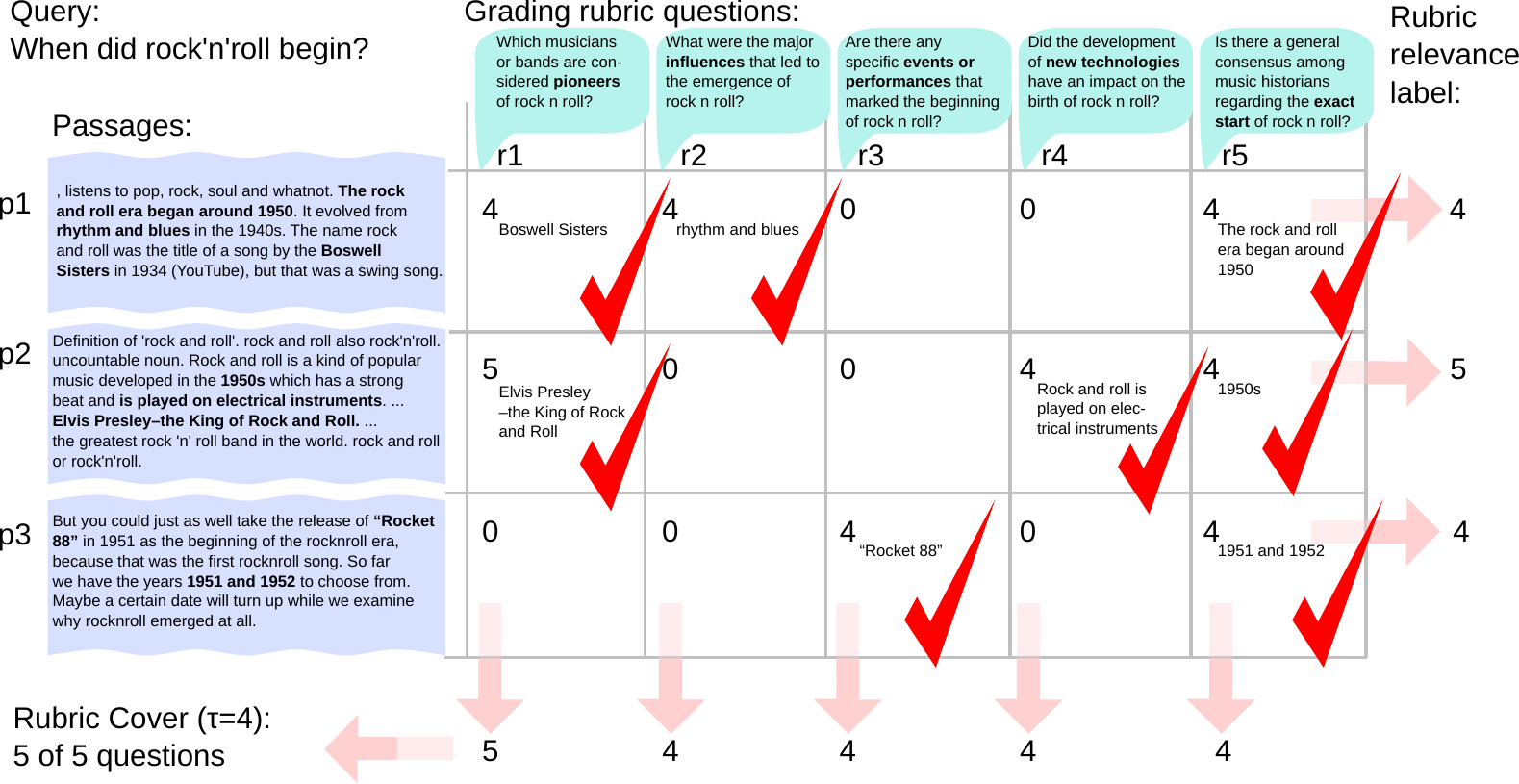}
    \caption{ \Ourapproach{} uses LLMs to grade how well a passage $p$ addresses each rubric question $r$.
    For each passage and question, the grade on a scale from 0 (worst) to 5 (best) is depicted in each cell. Extracted answers for verification.\\
    % Each cell depicts the grade assigned for each passage and question on a scale from 0 (worst) to 5 (best), along with extracted answers for manual verification.\\
    Relevance labels and coverage scores for the \EXAM{} evaluation are derived from these grades. This example is based on actual \EXAM{} grades obtained with our system for \dlsecond{} query \texttt{940547}.     \label{fig:approach}}
\fi
\end{figure*}

\section{Related Work}
\label{relatedwork}

Our work is unique in that it does not require any passage-level relevance judgments. We build on research ideas that we detail below.

\subsection{Traditional IR Evaluation}

The dominating approach to IR evaluation is based on (1) developing a collection of queries and a corpus, (2) collecting rankings from a range of retrieval systems, (3) building a judgment pool based on the top $k$ of each ranking, (4) hiring human judges to assess whether each retrieved passage (or document) is relevant or not---optionally using a graded relevance scale.

The evaluation process is based on simplifying assumptions of the Cranfield paradigm \cite{voorhees2009come}, such as that the relevance of each passage can be judged independently of other entries in the ranking.
This process is supported with several tools, such as \treceval{},\footnote{The \treceval{} tool is available at \url{https://github.com/usnistgov/trec_eval}} that determine the quality of a ranking based on the number of relevant passages (or documents) placed at high ranks. 
% Popular evaluation measures for ranking are normalized cumulative discounted gain (NDCG), (mean-)average precision (MAP), Precision@10, or set-based measures such as F1-measure, Area under the Receiver-Operator curve (ROC-AUC), or Utility.  
All these evaluation measures require that the relevance of each passage is judged, as ``holes'' can negatively affect the reliability of the evaluation \cite{macavaney2023one}.

This paradigm yields reusable test collections when passages from a provided corpus are ranked  without modification to the text. In this situation, we can expect that all relevant passages have been manually inspected, and hence any unjudged passages can be safely assumed to be non-relevant.  

However, when the assumption that unjudged passages are non-relevant no longer holds, the test collection is not reusable.
Violations of this assumption can be due to (1) increased corpus size, (2) reduced variability of ranking methods contributed to the judgment pool, and (3) limited size of judgment pools. Issues arise already when passages are extracted from document fragments \cite{allan2004hard}. The assumption no longer holds when text is generated with natural language generation or abstractive summarization---especially when the same system may produce a (slightly) different response every time. \textsl{Hence, with \thesystem{}, we offer an alternative evaluation paradigm for IR for the era of \acp{LLM}.}

\subsection{Nugget-based Evaluation}

In contrast to the Cranfield paradigm, work on max marginal relevance \cite{carbonell1998use} and search result diversification \cite{santos2015search} argue that redundancy should be avoided. This is captured by evaluation measures such as $\alpha$-NDCG \cite{sakai2017alphandcg,clarke2008novelty}, which define system quality by the coverage of a set of different query interpretations in the ranking.

Nugget-based evaluation is a concrete instance of this idea: Human judges identify important key facts (so-called ``nuggets''). The system quality is quantified by the number of key facts mentioned in the system's ranking. 
This idea is explored in the IR community for a ranking of passages/documents    and in the summarization community where a single text is to be assessed \cite{lin2006will,pavlu2012ir}.
The set of important nuggets is defined by a human judge during topic development and refined during the assessment process.  The common definition of a nugget is ``the smallest portion of text that constitutes relevant information in and of itself'' \cite{pavlu2012ir}.

Manually identifying which of the key facts/nuggets are mentioned in the systems' response requires additional manual work that poses a hurdle in the application of this evaluation approach. Unfortunately, this work needs to be repeated whenever a new system is to be evaluated. 

\sloppypar \textsl{\Oursystem{} supports evaluation with nuggets, while automatically matching nuggets mentioned in text with \acp{LLM}. The success of Chain-of-Thought prompting \cite{wei2022chain} leads to instruc\-tion-tuned \acp{LLM} that are trained to identify such nuggets.}

% There is a long history in the IR community to evaluate the relevance of documents by breaking down the information need into a set of ``nuggets'' (also called query intents, facts, or SCUs) that can each be evaluated independently \citep{lin2006will}. The common definition of a nugget is ``the smallest portion of text that constitutes relevant information in and of itself'' \cite{pavlu2012ir}. While text is defined as relevant as soon as it matches any nugget, the nugget document score is based on the sum of shingle match scores in any document---to be considered as relevant when surpassing a threshold.

% % \url{http://idl.iscram.org/files/mccreadie/2023/2529_McCreadie+Buntain2023.pdf}
% With the advent of LLMs, nugget-style evaluation is being revamped, most recently in the TREC Crisis Facts track \cite{mccreadie2023crisisfacts}: Judges are asked identify atomic ``facts'' (similar to nuggets). System responses are analysed for mentions of these facts, either via a boolean OR or with an embedding-based method.

% Our proposal is related in that our questions are intended to capture nuggets. A hurdle in using nugget-based evaluation method is to reliably match nuggets to text content. In contrast, with FLAN-T5-EXAM, we are leaning on decades-long research of factoid \questionanswering{}---which is now considered a solved task.

\subsection{Question-based Evaluation}\label{sec:rw-test-questions}

While the technology of matching nuggets in text is still new, several decades of research have been invested in the development of \questionanswering{} systems \citep{khashabi2020unifiedqa}.
 By designing a test question for each key fact, we can scan the IR system's response for possible answers. This evaluation is reminiscent of conducting an exam in school: the more exam questions can be answered based on information provided in the IR system's response, the higher the quality of the system. 

Originally attempted with manual answer verification \cite{doddington2002automatic}, it empirically showed success when an automatic \questionanswering{} system is used for identifying answers to test questions in the system's response \citep{farzi2024exam,sander2021exam}. The approach also demonstrated merit for the evaluation of summarization systems, when generating  multiple-choice questions \cite{Huang2020qasummary}, free text questions with exact-match answer verification \cite{deutsch2020questionanswering}, or entity-based questions \cite{wang-etal-2020-factual-qa,eyal-etal-2019-question-summary-first} from a gold summary or source text. 

% Recently, it was demonstrated that powerful \acp{LLM} are capable of generating good exam questions for IR system evaluation \citep{farzi2024exam}.

\textsl{\Ourapproach{} leans on the ability of \acp{LLM} to generate and answer questions with context, as recently demonstrated \citep{farzi2024exam}. Our workbench allows to analyze the quality of exam questions and refine the test bank.}

\subsection{LLM-based Relevance Assessment}

Since the release of ChatGPT, several researchers suggested using \acp{LLM} to directly assess the relevance of retrieved/generated passages \citep{macavaney2023one,faggioli2023perspectives,thomas2023large,liang2022holistic}. This approach uses a simple \ac{LLM} prompt such as ``does the passage answer the query?'' and uses the response as a relevance label. This approach mimics the passage-level relevance judgment as conducted with human judges, albeit with full automation. Empirically this approach is extremely successful, being able to reproduce the official leaderboard of the TREC Deep Learning track.
Similar prompts can be used for passage ranking \cite{sun2023chatgpt} or to evaluate the quality of \acp{LLM} \cite{murakhovska-etal-2022-mixqg}. 

BertScore \citep{zhang2019bertscore} and GptScore \citep{fu2023gptscore} work similarly but measure quality by the embedding similarity against a known good passage.

\textsl{We support many such methods for comparison.}

\subsection{LLM Evaluation with Human-in-the-loop}

Despite the empirically successful approach, critiques have been raised about using LLM's for relevance labeling. \citet{faggioli2023perspectives,faggioli2024determines} elaborate a wide range of theoretical concerns, centered on questions of trustworthiness and reliability of LLMs especially as new model versions are released in the future. \citet{wang2023large} empirically demonstrates that LLMs exhibit unfair positional bias. \citet{liu2023llms} confirm that when using a particular \ac{LLM} model for evaluation, it will prefer systems that use the same \acp{LLM} model for retrieval and/or generation.

\citet{fok2023search} study general issues of human over-reliance and under-reliance on LLMs. They elaborate why rationales produced by LLMs for human verification do generally not lead to improvements. In the context of judging the veracity of statements with crowd workers, \citet{xu2024on} find that \acp{LLM} can present wrong rationales convincingly, which can mislead inattentive human judges.
An open research question is how to achieve ideal competence partitioning \cite{hancock2013task} between humans and \acp{LLM} to obtain an IR system evaluation paradigm that is reusable, cost-effective, and trustworthy \citep{upcoming-cacm}. 

\textsl{Nugget- and question-based evaluation paradigms allow for a natural integration of a human-verifier-in-the-loop. We provide \thesystem{} as a proofing ground for developing novel approaches for integrating humans and LLMs into evaluation paradigms.}

\section{\system{} Approach}
\label{sec:approach}

\begin{table*}
\caption{ChatGPT Prompts for test bank generation. The instruction to respond in valid JSON is appended.}
\label{tab:generation-prompt}
\begin{tabular*}{1\linewidth}{@{\extracolsep{\fill}}@{\extracolsep{\fill}}>{\raggedright}p{0.1\linewidth}>{\raggedright}p{0.40\linewidth}>{\raggedright}p{0.40\linewidth}}
\toprule 
\textbf{Test Bank} & \textbf{Exam Question } & \textbf{Nugget / Key Fact}\tabularnewline
\midrule 
\textbf{\dl{}} & Break the query '\{query\_text\}' into concise questions that must
be answered. Generate 10 concise insightful questions that reveal
whether information relevant for '\{query\_text\}' was provided, showcasing
a deep understanding of the subject matter. Avoid basic or introductory-level
inquiries. Keep the questions short. \{instruction\} & Break the query '\{query\_text\}' into concise nuggets that must be
mentioned. Generate 10 concise insightful nuggets that reveal whether
information relevant for '\{query\_text\}' was provided, showcasing
a deep understanding of the subject matter. Avoid basic or introductory-level
nuggets. Keep nuggets to a maximum of 4 words. \{instruction\}\tabularnewline
\midrule 
\textbf{\car{}} & Explore the connection between '\{query\_title\}' with a specific
focus on the subtopic '\{query\_subtopic\}'. Generate insightful questions
that delve into advanced aspects of '\{query\_subtopic\}', showcasing
a deep understanding of the subject matter. Avoid basic or introductory-level
inquiries. \{instruction\} & Explore the connection between '\{query\_title\}' with a specific
focus on the subtopic '\{query\_subtopic\}'. Generate insightful nuggets
(key facts) that delve into advanced aspects of '\{query\_subtopic\}',
showcasing a deep understanding of the subject matter. Avoid basic
or introductory-level nuggets. Keep nuggets to a maximum of 4 words.
\{instruction\}\tabularnewline
\midrule 
\textbf{Instruction} & Give the question set in the following JSON format: 
\begin{verbatim}
```json 
{ "questions" : [question_text_1,
question_text_2, ...]
``` \end{verbatim}
& Give the nugget set in the following JSON format: 
\begin{verbatim}
```json 
{ "nuggets" : [nugget_text_1, nugget_text_2, 
...]} 
```  \end{verbatim}\vspace{-1em}\tabularnewline
\bottomrule
\end{tabular*}
\bigskip
\end{table*}

\begin{table*}
\caption{FLAN-T5 grading prompts for grading system responses (suggested by ChatGPT 4). Prompt class given in small font.}
\label{tab:prompt-table}
\begin{tabular*}{1\linewidth}{@{\extracolsep{\fill}}@{\extracolsep{\fill}}>{\raggedright}p{0.1\linewidth}>{\raggedright}p{0.40\linewidth}>{\raggedright}p{0.40\linewidth}}
\toprule 
\textbf{Grading } & \textbf{Exam Question } & \textbf{Nugget / Key Fact}\tabularnewline
\midrule 
\textbf{Self-Rated } & {  \footnotesize{\texttt{QuestionSelfRatedUnanswerablePromptWithChoices:} } }

Can the question be answered based on the available context? choose
one:

- 5: The answer is highly relevant, complete, and accurate.

- 4: The answer is mostly relevant and complete but may have minor
gaps or inaccuracies.

- 3: The answer is partially relevant and complete, with noticeable
gaps or inaccuracies.

- 2: The answer has limited relevance and completeness, with significant
gaps or inaccuracies.

- 1: The answer is minimally relevant or complete, with substantial
shortcomings.

- 0: The answer is not relevant or complete at all.

Question: \{question\}

Context: \{context\}  & 
{\footnotesize{\texttt{NuggetSelfRatedPrompt:} } }

Given the context, evaluate the coverage of the specified key fact
(nugget). Use this scale:

- 5: Detailed, clear coverage.

- 4: Sufficient coverage, minor omissions.

- 3: Mentioned, some inaccuracies or lacks detail.

- 2: Briefly mentioned, significant omissions or inaccuracies.

- 1: Minimally mentioned, largely inaccurate.

- 0: Not mentioned at all.

Key Fact: \{nugget\}

Context: \{context\}\tabularnewline
\midrule 
\textbf{Answer Extraction } & 
{\footnotesize{\texttt{QuestionCompleteConciseUnanswerablePromptWithChoices:} } }

provide a complete and concise answer to the question based on the
context.

Question: \{question\}

Context: \{context\}  & 
{\footnotesize{\texttt{NuggetExtractionPrompt:} } }

Extract the passage from the text that best relates to the key fact
(nugget), ensuring relevance and clarity.

Key Fact: \{nugget\}

Context: \{context\}\tabularnewline
\bottomrule
\end{tabular*}
\end{table*}

\textbf{Task statement.}
Given a set of queries and IR systems, we implement a workbench that aids with the following \textbf{evaluation task}:

An information retrieval/generation system is given a search \emph{query} $q$ to produce a relevant \emph{system response} comprised of passages $P$. The response can take the form of a passage ranking, a set of extractive summaries, or a single (generated) text. 
\\
Given system responses across queries from multiple systems, the task is to assign each system an \emph{evaluation score} that represents the quality of the information content provided by the system.

\medskip

\noindent \Ourapproach{}  proceeds as follows (cf.\ Figure \ref{fig:approach}).

\subsection{Phase 1: Test Bank Generation}

Human judges develop a test bank of nuggets and/or exam questions. To support this activity, our workbench can generate an initial set of test nuggets or exam questions with the help of ChatGPT, using one of the prompts given in Table \ref{tab:generation-prompt}. The human's role is to confirm and refine the test bank, by editing, adding, or removing nuggets or questions. Our process permits to refine the test bank at any time, for example by iterating the process based on outputs of Phase 3.

\subsection{Phase 2: Grading}

The \ac{LLM} will grade all system responses, passage-by-passage, by either scanning the passage for mentions of each nugget or trying to answer each exam question based on the passage content. \\ Our workbench supports three grading modes: \begin{description}
        \item[Self-rating answerability:] The \ac{LLM} is prompted to rate to which extent the passage contains information pertaining to the nugget or question on a scale from 0 (worst) to 5 (best).
        \item[Answer extraction with verification:] For the case of exam questions with a known correct answer, the \ac{LLM} is trying to answer the exam question with the passage content. Subsequently the produced answer is verified against the known correct answer and counted as ``correctly answered'' if the answers match.
        \item[Informational answer extraction:] To support human verification even without known correct answers, the \ac{LLM} is instructed to extract the answer to an exam question or extract the text span that mentions a nugget.
    \end{description}

\subsection{Phase 3: Manual Oversight and Verification}
    To manually verify that the \ac{LLM} is operating as intended, extracted answers and nugget  mentions should be inspected and cross-correlated with derived passage-level grades. Grading prompts may need to be adjusted in response to incorrectly graded passages. Known correct answers to exam questions may be manually expanded whenever correct answers are missed. 
    
    We recommend analyzing the effect of individual test bank entries on system rankings (leaderboards). Whenever manual relevance assessments are available, the verification can be complemented with additional analyses such as inter-annotator agreement and rank correlations with official leaderboards. 

    We envision that the test bank created in Phase 1 will be refined based on the inspection of graded results. For example, additional nuggets and exam questions should be created whenever relevant passages are missed by \oursystem{}.

\subsection{Phase 4: Evaluation}

An evaluation score for each IR system is computed based on the following assumption: the more nuggets are mentioned in, or the more questions can be answered with a system's response, the higher the system's evaluation score. Our workbench supports two evaluation measures: \begin{description}
        \item[\examcover{}:] A recall-based evaluation score computed as the fraction of test nuggets or exam questions covered with $k$ passages of the system's response.
        \item[\examqrels{}:] Exporting relevance labels, based on the grading level of at least one nugget or exam question. This metric is implemented by using the labels as a ``qrels'' file for \treceval{}, and hence supports a wide range of measures.
    \end{description}

\subsection{Direct Grading}
For comparison, we also implement direct grading prompts \citep{thomas2023large,faggioli2023perspectives,sun2023chatgpt,liang2022holistic}. These skip over phase 1 and yield passage-level relevance labels for evaluation with \treceval{}.

\section{Resource: \system{} Workbench} \label{sec:resource}

With this work we release \thesystem{} software, an implementation in python, along with example data. The different phases of \theapproach{} are implemented as individual commands so it is easy to build complex custom pipelines with a few lines in a python or bash script. Compressed JSON-lines is used as an interchange format, allowing to perform the grading phase on a high-performance GPU server, while human verification can be performed on a local computer with GUI support. Complex queries on data in compressed JSON-lines format can be performed with many available tools such as \texttt{jq},\footnote{Command line tool \texttt{jq} available at \url{https://jqlang.github.io/jq/}} and \texttt{DuckDB},\footnote{The data base DuckDB is available at \url{https://duckdb.org}} or directly in \texttt{python} via provided \texttt{pydantic}\footnote{Documentation for the \texttt{pydantic} package is available at \url{https://docs.pydantic.dev}}  bindings.

Installation and usage instructions of \thesystem{} is described in detail online.\footnote{\system{} software: \url{ \resourceurl }} All commands provide extensive line documentation when called with \texttt{--help}. 

Below we focus on describing the data model and elaborate steps for manual oversight and verification. In Section \ref{sec:walkthrough}, we provide a detailed walk-through using an example evaluation for the \dlsecond{} track.

\begin{figure}
\begin{footnotesize}
\begin{lstlisting}[basicstyle=\ttfamily,columns=fullflexible,  escapeinside={(*@}{@*)}]
{ "query_id": "940547",
  "query_text": (*@\colorbox{myyellow}{"when did rock n roll begin?"}@*),
  "info": {   "prompt_target": "questions"  },
 "items": [{
   "query_id": "940547",
   "question_id": "940547/a4c82219840e6d197d185ed1eda27c61",
   "question_text": (*@\colorbox{myyellow}{"Which musicians or bands are considered }@*)  
                        (*@\colorbox{myyellow}{pioneers of rock n roll?"}@*)
  },...
\end{lstlisting}
  % }, {
  %  "query_id": "940547",
  %  "question_id": "940547/851c0ef6dc72d20cb149576267d542af",
  %  "question_text": (*@\colorbox{myyellow}{"What were the major influences that led to the}@*) 
  %                      (*@\colorbox{myyellow}{emergence of rock n roll?"}@*)

\end{footnotesize}
\caption{Example question generated for \dlsecond{}.}
\label{fig:genq}
\end{figure}

\begin{figure}
\begin{footnotesize}
\begin{lstlisting}[basicstyle=\ttfamily,columns=fullflexible,  escapeinside={(*@}{@*)}]
{ "query_id": "940547",
  "query_text": (*@\colorbox{myyellow}{"when did rock n roll begin?"}@*),
  "info": { "prompt_target": "nuggets" },
  "items": [{
   "query_id": "940547",
   "nugget_id": "940547/3e9afdb8aeb54b6f496bb72040d7f212",
   "nugget_text": (*@\colorbox{myyellow}{"Early 1950s innovation"}@*)
  },...
\end{lstlisting}
  % }, {
  %  "query_id": "940547",
  %  "nugget_id": "940547/c3d5717ee691f3ae49cffb386986408f",
  %  "nugget_text": (*@\colorbox{myyellow}{"Rooted in blues"}@*)
\end{footnotesize}
\caption{Example nugget generated for \dlsecond{}.}
\label{fig:gen-nuggets}
\end{figure}

\subsection{Phase 1: Test Bank}

We envision that a test bank of nuggets or exam questions will be created with a semi-automatic approach.  In the example of \car{}, exam questions are directly available in the \tqa{} dataset. Test collections with manual nuggets annotations \cite{soboroff2016bolt} can be imported.
\Oursystem{} ingests exam questions and nuggets in JSON format depicted in Figures \ref{fig:genq} and \ref{fig:gen-nuggets}. 

To help seed the test bank creation process, \thesystem{} can generate exam questions and nuggets with the help of ChatGPT. The prompts for test bank generation are given in Table \ref{tab:generation-prompt}.
Examples of questions and nuggets generated for \dlsecond{} query ``940547'' are provided in  Figures \ref{fig:genq} and \ref{fig:gen-nuggets}. The question and nugget ID should be unique, as this key will be used when grading system responses. Our implementation uses an MD5-hash of the question text along with the query ID.

While we obtain strong empirical results with fully automatically generated questions and nuggets, the concerns raised by \citet{faggioli2023perspectives,faggioli2024determines} would still apply. We emphasize that a human-in-the-loop should manually verify the utility of the test bank, to ensure that test nuggets and questions are relevant for the query and are not missing subtle aspects that would indicate relevance. This manual refinement step can be performed before grading or during inspection of graded passages during manual oversight and verification in Phase 3. 

A potential concern is that systems under evaluation might be trained to merely address test questions and nuggets,  without intending to provide useful information to the user. To discourage systems that ``study to the test'', we suggest keeping parts of the test bank secret, and only sharing leaderboards and relevance labels.

\begin{figure}
\begin{footnotesize}
\begin{lstlisting}[basicstyle=\ttfamily,columns=fullflexible,  escapeinside={(*@}{@*)}]
[ (*@\colorbox{mygray}{"Query ID",}@*)
 [ {(*@\colorbox{mygray}{ "paragraph\_id": "Unique Paragraph Identifier",}@*)
    (*@\colorbox{mygray}{"text": "Full Text of the Paragraph",}@*)
    "paragraph": ... // additional markup if available.
    "paragraph_data": {
      "judgments": [ {
         "paragraphId": "Same Paragraph Identifier",
         "query": "Associated Query/Subquery ID",
         (*@\colorbox{myblue}{"relevance": 2,}@*) // judgment grade
         "titleQuery": "Query ID"
       }],
      "rankings": [ {
         (*@\colorbox{myblue}{"method": "Ranking Method or System Name",}@*)
         (*@\colorbox{myblue}{"paragraphId": "Unique Paragraph Identifier",}@*)
         "queryId": "Associated Query/Subquery ID",
         (*@\colorbox{myblue}{"rank": 6,}@*) // retrieval rank
         "score": 17.560528 // retrieval score
       }]
    },
    "exam_grades": [     // grades populated in Phase 2     
     { (*@\colorbox{lime}{"correctAnswered": ["Correctly Answered Question IDs"],}@*)
       (*@\colorbox{lime}{"wrongAnswered": ["Incorrectly Answered Question IDs"],}@*)
       (*@\colorbox{myyellow}{"self\_ratings"}@*): [{
            (*@\colorbox{myyellow}{"question\_id": "Question ID",}@*)
           // alternatively: "nugget_id": "Nugget ID"
           (*@\colorbox{myyellow}{"self\_rating": 4 // self-rating grade}@*)
         },
       (*@\colorbox{myorange}{"answers": [ ["Question ID", "Answer Text"] ],}@*)
       (*@\colorbox{mypurple}{"llm": "Huggingface Language Model Used",}@*)
       "prompt_info": {
          (*@\colorbox{mypurple}{"prompt\_class": "QuestionSelfRatedPrompt",}@*)
         "prompt_style": "Can the question be answered based on...",
         "context_first": false,
         "check_unanswerable": true,
         "check_answer_key": false,
         "is_self_rated": true
       },
       "exam_ratio": 0.25 // fraction of mentioned nuggets
     }],
    "grades": [ {  // for direct relevance grading
       (*@\colorbox{lime}{"correctAnswered": true,}@*) // if graded as relevant
       (*@\colorbox{myyellow}{"self\_ratings": 4}@*) // grade
       (*@\colorbox{myorange}{"answers": "Answer Text"}@*)
       (*@\colorbox{mypurple}{"llm": "Huggingface Language Model Used",}@*)
       (*@\colorbox{mypurple}{"prompt\_info" : {...},}@*)
     }]
  }]
]
\end{lstlisting}
       % "llm_options": { 
       %   % "prompt_template": "Template Used for Prompt",
       %   "answer_match": "Answer Matching Strategy"
       % },

\end{footnotesize}
    \caption{Data Model. \colorbox{mygray}{Query, passage text and ID}  must be provided. If available, \colorbox{myblue}{manual judgment level} and \colorbox{myblue}{system information} can be used for analysis and verification in Phase 3 and 4.  Phase 2 adds the fields \texttt{exam\_grades}  and/or \texttt{grades} with information about \colorbox{lime}{correct questions/nuggets},  \colorbox{myyellow}{self-ratings of answerability}, and \colorbox{myorange}{answers for manual verification}. %Phase 3, the workbench derives different views onto the data. 
    All phases support filtering based on fields \colorbox{mypurple}{\texttt{llm} and \texttt{prompt\_class}}.
    }
    \label{fig:datamodel}
\end{figure}

\subsection{Data Model for Autograding}
\label{sec:datamodel}

Figure \ref{fig:datamodel} depicts the JSON data model.  To use \thesystem{},  system responses are converted into our format, providing at a minimum \texttt{Query ID}, \texttt{paragraph\_id}, and \texttt{text} (depicted in light grey in Figure \ref{fig:datamodel}).  Optionally, the \texttt{paragraph} can be provided in full markup, if helpful for manual verification support. 

If applicable and available, manual judgments,  system information, and information of a ranking can be provided (cyan). If not available, \texttt{judgments} and \texttt{rankings} can be set to an empty list. For systems that are generating a longer response instead of a ranking, the field \texttt{rank} can be used to indicate a sequence or priority (or be set to ``1''). During Phase 3 (Evaluation) it is used as a cut-off criterion for \examcover{}.

Phase 2 (Grading) will populate the fields \texttt{exam\_grades} with information about the coverage of nuggets and answerability of exam questions. Direct grading prompts will populate the field \texttt{grades} (\texttt{Sun} and \texttt{Sun\_few}  \cite{sun2023chatgpt}, \texttt{Fag} and \texttt{Fag\_fewshot}  \cite{faggioli2023perspectives}, \texttt{Thom} \cite{thomas2023large}, and \texttt{HELM} \cite{liang2022holistic}). 
In both cases, \texttt{correctAnswered} will list the nuggets/questions correctly addressed (or true/false for direct grading). The field \texttt{wrongAnswered} lists nuggets/questions that the system response did not appropriately address.  In particular, this applies to prompts that verify the correctness of the grader LLM's response, as in the case for \questionanswering{} with a known correct answer or exact match of a nugget. 

We suggest to identify addressed questions and nuggets with self-rated prompts given in Table \ref{tab:prompt-table}. The grader-LLM will rate the extent to which the provided passage covers respective questions and nuggets on a scale from 0 to 5. This information is provided in field \texttt{self\_ratings} along with the \texttt{nugget\_id} or \texttt{question\_id}. 

For manual verification of the process in Phase 3, the raw answer of the grader-LLM is stored in the \texttt{answers} field, along with nugget or question ID. 

Information about the Hugging face model used as grader-LLM is stored in the field \texttt{llm}. The grading prompt class and options are preserved in the field \texttt{prompt\_info}, available for grade selection for the evaluation in Phase 4.

\subsection{Phase 2: Grader-LLMs and Prompts}
The grader-LLM can work with a wide range of models from Hugging face,\footnote{Models available at \url{https://huggingface.co/models}}  supporting the Hugging face pipelines \texttt{text2text-gen\-eration} (for the T5 family), \texttt{text-generation} (for OLMo, Llama-2, and gpt2 models), and \texttt{question-answering} (for models that are fine-tuned on \squad{}).

While the grading prompts can be adjusted by the user of \thesystem{}, we provide a set of default prompts for grading which exam questions are answerable and which nuggets are mentioned. The prompts in Table \ref{tab:prompt-table} are designed to work well with models of the FLAN-T5 family. They were chosen based on suggestions from GPT-4.

\subsection{Phase 3: Manual Verification and Oversight}

A key concern about using LLMs, especially via self-rating prompts, is how to diagnose when the LLM fails to correctly assess whether a question/nugget is addressed in a passage. 
If many low-quality passages are associated with high grades, the grading prompts need to be adjusted, or a stronger grading LLM needs to be chosen.

To ensure that as-relevant-rated passages are indeed relevant, we recommend that human verification should inspect passage text with a high rating along with the corresponding question/nugget and extracted answer. To reduce the passage pool, emphasis can be placed on passages with at least $m$ highly graded answers.

% We suggest to scan for passages which correctly address exam questions or nuggets,  that the grader-LLM fails to identify. To screen those cases, we list passages where the self-rating is low, but the extracted answer matches important key words.

\Oursystem{} offers support to verify that grading levels correlate with the quality of question's answers and extraction of nuggets via the following analyses:

    \begin{description}
        \item[Verify grading:] To confirm the quality of the grading-LLM, this analysis lists extracted answers and corresponding grades question-by-question. This test indicates when grading prompts need to be adjusted.
        \item[Grid display:] To oversee the process, passages with grades per test question/nugget along with extracted answers are listed.
        \item[Missing in test bank:] To grow the test bank as needed, this analysis lists relevant passages that do not address any of the current test nuggets or questions. The test bank needs to be extended to account for the relevant information in those passages.
        \item[Spurious test bank entries:] To identify spurious test bank entries, which would yield false positive relevance labels, this analysis lists test nuggets or questions that are frequently covered by non-relevant passages. Those entries should be removed from the test bank or be reformulated.
    \end{description}

% \bigskip

\noindent Additional analyses are easy to derive by tools that handle JSON formatted files, such as \texttt{DuckDB} or \texttt{jq}.

\begin{table*}[ht]
\caption{Description of filenames for \dlsecond{} walk through, available in the online appendix.}
\label{tab:filenames-dl20}

\begin{tabular*}{1\linewidth}{>{\raggedright}p{0.08\linewidth} >{\raggedright}p{0.21\linewidth} >{\raggedright\arraybackslash}p{0.65\linewidth}}
\toprule 
\textbf{Phase} & \textbf{Filename} & \textbf{Description}\tabularnewline
\midrule 
\multirow{3}{*}{\makecell{\textbf{External} \\ \textbf{Input}}} & \colorbox{mylight}{\texttt{dl20-queries.json}} & JSON dictionary mapping query ID to query text.\tabularnewline
 & \multirow{2}{*}{\colorbox{mylight}{\texttt{trecDL2020-qrels-runs-with-text.jsonl.gz}}} \tabularnewline
 &  & \vspace{1pt}Passages selected for grading, providing Query ID, paragraph\_id,
text, paragraph\_data as described in Figure \ref{fig:datamodel}.\tabularnewline
\midrule 
\multirow{2}{*}{\makecell{\textbf{Phase 1:} \\ \textbf{Test Bank}}}& \colorbox{mylight}{\texttt{dl20-questions.jsonl.gz}} & Generated exam questions\tabularnewline
 & \colorbox{mylight}{\texttt{dl20-nuggets.jsonl.gz}} & Generated test nuggets\tabularnewline
\midrule 
\multirow{2}{*}{\makecell{\textbf{Phase 2:} \\ \textbf{Grading}}}& \texttt{\commentout{\texttt{questions-explain-{}-questions-rate-{}-nuggets-explain-{}-nuggets-rate-{}-all-trecDL2020-qrels-runs-with-text.jsonl.gz}}\colorbox{mylight}{\texttt{questions-explain-{}-questions-rate-{}-nuggets-...trecDL2020-qrels-runs-with-text.jsonl.gz} }} & \tabularnewline
 &  & Passages graded with nuggets and questions using the self-rating prompt
(for formal grades) and the answer extraction prompt for manual verification.
Grade information is provided in the field exam\_grades.\tabularnewline
\midrule 
\multirow{3}{*}{\makecell{\textbf{Phase 3:} \\\textbf{Manual} 
\\\textbf{Verification}}}& \colorbox{mylight}{\texttt{dl20-verify-grading.txt}} & All LLM responses are grouped by test question/nugget.
by question/nugget ID.\tabularnewline
 & \colorbox{mylight}{\texttt{dl20-bad-question.txt}} & Questions/Nuggets that frequently obtain a grade above 4, judgment below 1.\tabularnewline
 & \colorbox{mylight}{\texttt{dl20-uncovered-passages.txt}} & \hspace{1.5em}Passages judged relevant without any grade above 4.\tabularnewline
 & \colorbox{mylight}{\texttt{dl20-grid-display.txt}} & Passages judged relevant without any grade above 4.\tabularnewline
\midrule 
\multirow{3}{*}{\makecell{\textbf{Phase 4:} \\ \textbf{Evaluation}}}& \colorbox{mylight}{\texttt{dl20-\systemfile{}-qrels-\$promptclass-minrating-4.solo.qrels}} & \tabularnewline
 &  & Exported Qrel file treating passages with grades $\geq4$ as
relevant.\tabularnewline
 & \colorbox{mylight}{\texttt{dl20-\systemfile{}-qrels-leaderboard-\$promptclass-minrating-4.solo.mrr.tsv}} & \tabularnewline
 &  & Leaderboard produced with \treceval{} using the Qrel file above
and metric MRR.\tabularnewline
 & \colorbox{mylight}{\texttt{dl20-\systemfile{}-cover-leaderboard-\$promptclass-minrating-4.solo.tsv}} & \tabularnewline
 &  & Leaderboads produced with \examcover{} treating test nuggets/questions
as answered when any passage obtains a grade $\geq4$.\tabularnewline
\midrule 
\makecell{\textbf{Analyses}\\ \hfill{}\textbf{ Input:}}& \colorbox{mylight}{\texttt{official\_dl20\_leaderboard.json}} & \hspace{3em} JSON dictionary with method names mapped to leaderboard ranks (top
rank = 1).\tabularnewline
\textbf{\textbf{\hfill{}} Output:} & \colorbox{mylight}{\texttt{dl20-\systemfile{}-qrels-leaderboard-\$promptclass-minlevel-4.correlation.tsv}} & \tabularnewline
 &  & \examqrels{} leaderboard as before, but including rank correlation
with the official DL20 leaderboard, using Spearman's rank and Kendall's
tau correlations. (*) \tabularnewline
 & \colorbox{mylight}{\texttt{dl20-\systemfile{}-cover-leaderboard-\$promptclass-minlevel-4.correlation.tsv}} & \tabularnewline
 &  & \examcover{} leaderboard as before with rank correlation with the
official leaderboard. (*) \tabularnewline
 & \colorbox{mylight}{\texttt{dl20-\systemfile{}-inter-annotator-\$promptclass.tex}} & \tabularnewline
 &  & LaTeX tables with graded and binarized inter-annotator statistics
with Cohen's kappa agreement. ``Min-answers'' refers to the number
of correct answers obtained above a grading threshold by a passage. (*)\tabularnewline
&\multicolumn{2}{l}{(*) For \dl{} \texttt{\textendash -min-relevant-judgment 2} must be
set.} \tabularnewline
\bottomrule
\end{tabular*}
\end{table*}

\subsection{Evaluation Metrics}

After refining the test bank and verifying that the grading process works as expected, evaluation scores for each graded system can be exported via two mechanisms:  \examqrels{} and \examcover{}.

For \examqrels{}, a \treceval{}-compatible relevance file is exported, where grades are aggregated into a passage-level relevance label. Our current implementation can be configured to derive the passage relevance label as either (1) number of correctly covered test nuggets/questions, or (2) the highest grades across  all test nuggets/questions. Additional measures are easy to implement in our code base. The exported relevance labels (``qrels'') can be shared with system developers, so they can evaluate their systems with  \treceval{}.  Furthermore, our implementation supports to automatically obtain evaluation measures from \treceval{} on a collection of ``run''-files, to produce a leaderboard of system evaluation scores. %\footnote{Our implementation currently does not support error bars or significance tests for \examqrels{}, but we encourage adopters to use appropriate tooling for such tests.}

For classes of information needs that rely on covering all relevant information, we provide the \examcover{} measure.
Rather than obtaining a passage-level relevance label, we consider the total number of test nuggets/questions addressed across the first $k$ passages $p$ of a system's response $P$. This applies to system responses in the form of a passage ranking as well as to multi-passage responses produced with natural language generation methods. Our implementation will assign a quality score for each system based on the fraction of test nuggets or questions that are addressed. For grading prompts with self-rating, our implementation supports different grade thresholds, at which a test nugget/question is considered covered.

We imagine that several similar evaluation measures can be implemented on the basis of these ideas, including $\alpha$-NDCG \cite{sakai2017alphandcg,clarke2008novelty}.

\section{Code Release}

We release the Autograder Workbench as a python implementation along with some example data. The code is released under the BSD-3 open-source license. The workbench is implemented as a Python library which also offers a command-line interface for each phase when installed with \texttt{poetry}. To support reproducibility, we lean on the \texttt{nix} build system for releases of \thesystem{} software.

\section{Walk-through on \dlsecond{}}
\label{sec:walkthrough}

We provide a resource walk-though on an example application to \dlsecond{} \cite{dl20}. Data is provided as a separate archive,\footnote{\system{} files for \dlsecond{} available in the online appendix.} to be unpacked into directory \texttt{./data/dl20/}.

%\paragraph{Configuration.}
For question generation, we use \texttt{gpt-3.5-turbo-} \texttt{instruct}; for grading we use \texttt{google/} \texttt{FLAN-T5-lar\-ge} with the \texttt{text2text-gen\-eration} pipeline from Hugging Face.\footnote{\url{https://huggingface.co/google/flan-t5-large}} In general, \texttt{FLAN-T5-large} provides a good trade-off between quality in responses and computational cost. We are able to run all computational steps of this walk-through in 8 hours, using a API access to OpenAI for Phase 1 and a single NVIDIA A40 GPU for Phase 2.

Table \ref{tab:filenames-dl20} gives an overview of all files used as input/output for this walkthrough of the \system{} process.

\textbf{External input:} Query IDs and query text are converted to a JSON dictionary.
We select all passages in official judgments and the top 20 of all run submissions---a total of 11,386 passages. For each passage we provide Query ID, paragraph\_id, text, paragraph\_data as gzip compressed JSON-lines following the format described in Figure \ref{fig:datamodel}. 

To compare to a generative system, we include several GPT-based systems that did not contribute to the judgment pool. We segment the response into paragraph-size passages of about 400 words.

%Our code provides python data objects that for \texttt{pydantic} to simplify this process.
\textbf{Phase 1:}
With our test bank generation prompts, we obtain roughly ten exam questions and test nuggets for each of the 54 queries in \dlsecond{}. Examples are depicted in Figures \ref{fig:genq} and \ref{fig:gen-nuggets}. A description of file names is given in Table \ref{tab:filenames-dl20}.

\textbf{Phase 2:}
We grade all passages according to each test bank entry using the grading prompts in Table \ref{tab:prompt-table}. Since our generated test bank is comprised of open-ended questions, the answer extraction prompts are merely used for manual verification. 

\textbf{Phase 3:} 
To support a human supervising the process,\footnote{Detailed data available in the online appendix.} we first \textbf{verify  the grading}. Below examples for the question ``Which musicians or bands are considered pioneers of rock n roll?''
\begin{itemize}
    \item (rating: 5) Elvis Presley–the King of Rock and Roll \checkmark (true pos.)
    \item (rating: 0) rhythm and blues \checkmark (true neg.)
    \item (rating: 0) the 1930s \checkmark (true neg.)
\end{itemize}
% To analyze that a high self-rating is given when a good answer can be extracted, we obtain the answers of the grader-LLM across all passages grouped by question\_id (or nugget\_id). 

\noindent Next, we identify \textbf{spurious test bank entries}, i.e., questions and nuggets, that are often covered by non-relevant passages. For the example query \texttt{940547} on rock and roll (frequency in parentheses):
\begin{itemize}
\item (116)    Did rock n roll start as a distinct genre or did it evolve from existing music styles?
\item (102)    Is there a general consensus among music historians regarding the exact start of rock n roll?
\item (60)     Were there any notable recordings or songs that played a significant role in popularizing rock n roll?
\end{itemize}
% 940547/ed1f3ad87ff18418e629eaf306799441  116    Did rock n roll start as a distinct genre or did it evolve from existing music styles?
% 940547/1a9b463d18827c22e5f7e3a9b1f56364  102    Is there a general consensus among music historians regarding the exact start of rock n roll?
% 940547/6971da9b44d774dd31dbc3f8f4f48f39  60     Were there any notable recordings or songs that played a significant role in popularizing rock n roll?

\noindent To detect when the \textbf{test bank is missing important entries} and needs to be augmented, we list relevant passages that do not have any high grades. Such passages do not exist for this example query. 
However, we find such passages for query \texttt{1108651} ``what is the best way to get clothes white'':
 \begin{quote}
Use bleach. If you are only washing white clothes, add a capful of bleach to the load when you plan to wash it. If you are concerned about the powerful effects of standard bleach, try a non-chlorine bleach or a slow-acting bleach, like a 3-percent peroxide solution.
 \end{quote}

This passage does not answer the related rubric question ``How does soaking clothes in bleach affect their whiteness?'' It should be reformulated as ``How to use bleach to wash white clothes?''

\textbf{Phase 4:}
Table \ref{tab:dl20-qrels-leaderboard} displays the leaderboard generated with \examqrels{}, using different grade cutoffs. \Oursystem{} exports relevance labels as a ``qrels'' file, to evaluate with \treceval{}. Here, we use mean reciprocal rank (\texttt{MRR}), but any evaluation measure can be used with this approach. 
% For self-rated approaches, the minimum rating for passages to be relevant needs to be set. Our example uses 4, which represents  a ``strict'' binary relevance scale. 
%\Thesystem{} can automatically run \treceval{} on all run-files to capture the leaderboard.

The leaderboard obtained with \EXAM{} Question-4 correlates best with the official leaderboard, with comparable results achieved by \texttt{Fag} \cite{faggioli2023perspectives}. Question-based methods work well across different grading thresholds and \treceval{} metrics.
Nuggets are matched too frequently to discriminate between systems.

% by executing the following command and parsing the results:
% \begin{footnotesize}
% \begin{verbatim}
% for f in *.run; do  
%   res=`trec_eval -m P.20 -l 4 autograde.qrels $f`;
%   echo \"$f $res\"; 
% done
% \end{verbatim}
% \end{footnotesize}

% Preview source code for paragraph 4
\begin{table}
\caption{Leaderboard for \dlsecond{} using \examqrels{} to score by reciprocal rank on exam questions and test nuggets with different grade thresholds, sorted by question-4. Additional text generation baselines are included by us (in italics, denoted with *).    Bottom: Leaderboard correlation and direct grading prompts
from Sun \cite{sun2023chatgpt}, Fag + fewshot \cite{faggioli2023perspectives}, Thom \cite{thomas2023large}, and HELM \cite{liang2022holistic}.
}
\label{tab:dl20-qrels-leaderboard}
\begin{footnotesize}

\begin{tabular}{lccccccc}
\toprule 
System  & \multicolumn{3}{c}{Question} & \multicolumn{3}{c}{Nuggets} & Official\tabularnewline
\hspace{4em}grade$\geq$  & 3  & 4  & 5  & 3  & 4  & 5  & rank\tabularnewline
\cmidrule(lr){2-4}\cmidrule(lr){5-7}
\textit{GPT4-question{*}} & 0.827 & 0.747 & 0.658 & 1.000 & 1.000 & 0.982 &  \!\!\!\!$\approx 1$\tabularnewline
\textit{GPT3.5-question{*}} & 0.812 & 0.743 & 0.650 & 1.000 & 1.000 & 0.991 &  \!\!\!\!$\approx 1$\tabularnewline
pash\_f3 & 0.856 & 0.738 & 0.579 & 0.991 & 0.982 & 0.978 & 3 %\tabularnewline
\vspace{-0.3em}\tabularnewline\vspace{-0.3em} &  & &  & &  &  & $\dots$ \tabularnewline
bigIR-T5xp-T5-F & 0.723 & 0.630 & 0.531 & 0.991 & 0.982 & 0.969 & 27\tabularnewline
\textit{GPT3.5-wiki {*}} & 0.723 & 0.627 & 0.497 & 1.000 & 1.000 & 1.000 &  \!\!\!\!$\approx 27$\tabularnewline
TUW-TK-2Layer & 0.779 & 0.622 & 0.511 & 0.991 & 0.968 & 0.956 & 34
\vspace{-0.3em}\tabularnewline\vspace{-0.3em} &  & &  & &  &  & $\dots$ \tabularnewline
% \tabularnewline
% p\_d2q\_bm25rm3 & 0.730 & 0.591 & 0.459 & 0.991 & 0.991 & 0.974 & 36\tabularnewline
% p\_d2q\_bm25 & 0.705 & 0.589 & 0.456 & 0.982 & 0.958 & 0.949 & 35\tabularnewline
terrier-InL2 & 0.675 & 0.541 & 0.442 & 1.000 & 0.975 & 0.963 & 44\tabularnewline
\textit{GPT4-wiki {*}} & 0.663 & 0.528 & 0.418 & 1.000 & 1.000 & 0.991 &  \!\!\!\!$\approx 44$\tabularnewline
terrier-BM25 & 0.672 & 0.528 & 0.410 & 0.991 & 0.977 & 0.965 & 45
\vspace{-0.3em}\tabularnewline\vspace{-0.3em} &  & &  & &  &  & $\dots$ \tabularnewline
%\tabularnewline
% bcai\_class\_pass & 0.650 & 0.513 & 0.390 & 0.991 & 0.986 & 0.949 & 41\tabularnewline
% bm25\_bert\_token & 0.636 & 0.513 & 0.420 & 0.991 & 0.985 & 0.976 & 51\tabularnewline
% indri-fdm & 0.653 & 0.510 & 0.433 & 0.991 & 0.986 & 0.986 & 43\tabularnewline
% TF\_IDF\_d\_2\_t\_50 & 0.608 & 0.507 & 0.349 & 0.975 & 0.958 & 0.914 & 53\tabularnewline
\textit{GPT3.5-web {*}} & 0.597 & 0.506 & 0.466 & 0.932 & 0.932 & 0.932 & \!\!\!\!$\approx 49$%\tabularnewline
% p\_bm25rm3 & 0.603 & 0.504 & 0.378 & 0.970 & 0.956 & 0.943 & 49\tabularnewline
% indri-sdm & 0.626 & 0.498 & 0.425 & 0.955 & 0.951 & 0.951 & 48\tabularnewline
% bl\_bcai\_mdl1\_vt & 0.639 & 0.493 & 0.356 & 0.982 & 0.966 & 0.957 & 40\tabularnewline
% p\_bm25 & 0.617 & 0.482 & 0.397 & 0.972 & 0.965 & 0.961 & 50\tabularnewline
\vspace{-0.3em}\tabularnewline\vspace{-0.3em} &  & &  & &  &  & $\dots$ \tabularnewline
% indri-lmds & 0.614 & 0.477 & 0.417 & 1.000 & 0.985 & 0.975 & 47\tabularnewline
\textit{GPT4-web {*}} & 0.630 & 0.473 & 0.361 & 0.907 & 0.889 & 0.889 &  \!\!\!\!$\approx 47$%\tabularnewline
\vspace{-0.3em}\tabularnewline\vspace{-0.3em} &  & &  & &  &  & $\dots$ \tabularnewline
% terrier-DPH & 0.619 & 0.454 & 0.395 & 1.000 & 0.986 & 0.986 & 52\tabularnewline
% DLH\_d\_5\_t\_25 & 0.559 & 0.451 & 0.354 & 0.971 & 0.956 & 0.944 & 46\tabularnewline
% bl\_bcai\_mdl1\_vs & 0.618 & 0.433 & 0.358 & 0.991 & 0.986 & 0.958 & 42\tabularnewline
% small\_1k & 0.399 & 0.340 & 0.210 & 0.714 & 0.699 & 0.681 & 54\tabularnewline
% DoRA\_Large\_1k & 0.353 & 0.319 & 0.195 & 0.681 & 0.673 & 0.657 & 56\tabularnewline
% med\_1k & 0.363 & 0.313 & 0.183 & 0.701 & 0.693 & 0.664 & 55\tabularnewline
% DoRA\_Small & 0.234 & 0.155 & 0.128 & 0.743 & 0.678 & 0.664 & 57\tabularnewline
% DoRA\_Med & 0.189 & 0.130 & 0.103 & 0.734 & 0.682 & 0.642 & 58\tabularnewline
DoRA\_Large & 0.186 & 0.113 & 0.104 & 0.707 & 0.666 & 0.628 & 59\tabularnewline
\midrule 
% &  &  &  &  &  &  & \tabularnewline
Spearman  & 0.937 & \!\!\textbf{0.941}\!\!& 0.845 & 0.275 & 0.609 & 0.838 & \tabularnewline
Kendall  & 0.800 & \!\!\textbf{0.810}\!\! & 0.656 & 0.203 & 0.449 & 0.656 & \tabularnewline
\midrule 
``direct''  & Sun  & Thom  & Fag  & \multicolumn{2}{c}{Fag fewshot} & \!HELM\! & \tabularnewline
Spearman  & 0.924 & 0.751 & \!\!\textbf{0.940}\!\! & \multicolumn{2}{c}{0.918} & 0.930 & \tabularnewline
Kendall  & 0.785 & 0.623 & \!\!\textbf{0.815}\!\! & \multicolumn{2}{c}{0.786} & 0.785 & \tabularnewline
\bottomrule
\end{tabular}

\end{footnotesize}
\end{table}

\begin{table}
\caption{Inter-annotator agreement with manual judgments.}\label{tab:inter-annotator-DL20}
\begin{small}
\begin{tabu}{@{}llcclr@{}}%
\toprule%
\multirow{2}{*}{}&\textbf{Grade}&\multicolumn{2}{c}{\textbf{Judgments}}&\textbf{Total}&\textbf{Cohen's }$\boldsymbol{\kappa}$\\%
\cmidrule(l@{\tabcolsep}){3-4}%
&&2--3&0--1&&\\%
\cmidrule(l@{\tabcolsep}){1-6}%
\multirow{2}{*}{Question}&4--5&\fbox{\textbf{998}}&2377&3375&0.25\\%
&0--3&668&\fbox{\textbf{7343}}&8011&0.25\\\bottomrule%
\multirow{2}{*}{Nugget}&4--5&\fbox{\textbf{1211}}&4095&5306&0.16\\%
&0--3&455&\fbox{\textbf{5625}}&6080&0.16\\\bottomrule%
\end{tabu}%
\end{small}

\end{table}

When the qrels file is shared with system developers, new systems are likely to retrieve new passages which need to be graded with our workbench to avoid bias against ungraded passages \cite{macavaney2023one}. 

Alternatively, \thesystem{} provides an implementation of \examcover{}\texttt{@20}, which measures the fraction of the test bank correctly addressed with the overall system response.\footnote{Results provided in the online appendix.} % Our implementation uses the information provided in \texttt{paragraph\_data.rankings}. 
The leaderboard offers \examcover{} evaluation scores with standard errors, as well as a normalized \examcover{} score as suggested in earlier work \cite{sander2021exam}.

% As direct relevance prompts are not compatible with a coverage-based approach, these can only be  evaluated via the \examqrels{} approach.

\textbf{Evaluating text generation.}  We demonstrate that our approach can be applied to text generation approaches which were not submitted as systems to the \dlsecond{} track. We ask GPT-4 and 3.5 to generate a system response that imitates a Wikipedia article, a Web page, and gives direct answer for each query. The results are integrated in Table \ref{tab:dl20-qrels-leaderboard}, marked with ``*'', with approximate official ranks if sorted by Question-4. 

We demonstrate that \examqrels{} is able place these methods on the leaderboard. For example, GPT*-question approaches rank above all submitted systems, GPT4-wiki in the middle and GPT*-web ranks near the bottom of the leaderboard. We remark that since questions and nuggets are generated with GPT 3.5, the scores of GPT3.5 generative methods may be inflated.

% For example, asking GPT for a Wikipedia-like article on the query (\texttt{gpt4-wiki}) covers as many test questions (min rating 4) as the best submitted systems---but due to the much fewer passages, it obtains a poor \examqrels{} score. However asking GPT merely for an answer to the question, has the quality of the 56'th system on the official leaderboard.

\textbf{Post-hoc analyses.}
For \dl{}, an official leaderboard and manual judgments are available. These can be used to study the usefulness of \theapproach{}.
We provide abridged results in Table \ref{tab:dl20-qrels-leaderboard}. %\footnote{Results differ slightly from prior work \ld{update }\citep{farzi2024exam}, due to a different test bank.}
% Use of a different test bank than prior work \citep{farzi2024exam}, yielded slightly different results.} 
While our focus is to provide a resource to explore the merits of the \system{} evaluation paradigm, the high correlation with the official leaderboard demonstrates its usefulness.

Using the official judgments created by TREC assessors, we can study the inter-annotator agreement between the generated questions/nuggets and human judges. Example results are displayed in Table \ref{tab:inter-annotator-DL20}, noting that judgment 1 indicates non-relevance in \dl{}. While our goal was not to imitate the passage-level judgment process, we find a reasonable agreement with TREC judges (without tuning the  system). This agreement would likely improve when a human is integrated into the development of the test bank.
%Not very surprisingly, direct grading approaches can obtain a slightly higher $\kappa$ of up to \ld{verify }0.30,\footnote{Even higher results are obtained with ChatGPT 4.0 instead of \texttt{flan-t5-large}.} instead of the 0.25 obtained here. 
Direct grading approaches can obtain a higher $\kappa$ (especially with GPT 4). 
However, our emphasis is to identify a worthwhile alternative that avoids passage-level relevance assessment altogether and that lends itself to integrating a human-in-the-loop.

We envision that manual effort is focused on passages where manual judgments and \system{}'s relevance labels disagree. For example, many relevant passages only obtain self-rated grades of zero. This implies that additional test nuggets or questions need to be added to the test bank. Manual answer verification efforts should focus on passages that obtain a high grade but were explicitly judged as non-relevant.

\section{Conclusion}
\label{sec:conclusion}

We release the software for \thesystem{} (URL in abstract), a toolkit that allows to experiment with different approaches to incorporate LLMs into the evaluation process along with a human-in-the-loop. We focus on ideas that avoid passage-level relevance assessments, as these are difficult for humans to verify without being prone to over-reliance on LLMs or costly manual judgments. Concretely, we incorporate two ideas that were studied in IR literature before:  coverage of test nuggets \cite{lin2006will,pavlu2012ir} and exam questions \cite{sander2021exam,farzi2024exam}. In both cases, a grader-LLM is used to automate the cost-intensive parts of scanning system responses for mentions of key facts and answers. However, addressing concerns about the trustworthiness of LLMs \cite{faggioli2023perspectives}, we offer a workbench that supports the integration of manual supervision in the \system{} process. In particular, we are concerned about verifying the grading process (without performing passage-level relevance judgments manually) and providing an analysis framework to diagnose the coverage of the test bank. To instill trust in the process, we provide analyses to study correlation with official leaderboards and agreement with official manual judgments.

To ease the rate of adoption, we (1) support to initialize a test bank with automatically generated nuggets and questions and (2) offer an easy way to integrate with the evaluation tool \treceval{}.

% We hope that this resource will help to develop novel evaluation approaches for both information retrieval and retrieval-augmented generation systems.
We hope this resource will help develop novel evaluation approaches for information retrieval and retrieval-augmented generation systems.

\begin{acks}
This material is based upon work supported by the National Science Foundation under
    Grant No. 1846017. Any opinions, findings, and conclusions or recommendations expressed in this material
    are those of the author(s) and do not necessarily reflect the views of the National Science
    Foundation.
\end{acks}

\eject{}
\balance{}
\bibliographystyle{ACM-Reference-Format}
\bibliography{bibliography}

\end{document}